\documentclass[jpdc]{apjrnl}
\usepackage{graphics}
\begin{document}

\title{A Multilevel Approach to Topology-Aware Collective Operations 
        in Computational Grids}

\author{Nicholas T. Karonis
\\Department of Computer Science
\\Northern Illinois University
\\DeKalb, IL~~60115
\\Argonne National Laboratory
\\Argonne, IL~~60439
\\Email: karonis@niu.edu
\and         %
Bronis de Supinski
\\Center for Applied Scientific Computing
\\Lawrence Livermore National Laboratory
\\Livermore, CA~~94551
\\Email: bronis@llnl.gov
\and         
Ian Foster
\\Argonne National Laboratory
\\Argonne, IL~~60439
\\The University of Chicago
\\Chicago, IL~~60637
\\Email: foster@mcs.anl.gov
\and         
William Gropp and Ewing Lusk
\\Mathematics and Computer Science Division
\\Argonne National Laboratory
\\Argonne, IL~~60439
\\Email: gropp@mcs.anl.gov, lusk@mcs.anl.gov
\and         
Sebastien Lacour
\\IRISA / INRIA Rennes
\\University of Beaulieu
\\35042 Rennes, France
\\Email: Sebastien.Lacour@irisa.fr
}
\date{April 2002}

\maketitle

Proposed running head: Multilevel Topology-Aware Collective Operations

\pagebreak

\begin{abstract}
The efficient implementation of collective communication operations has 
received much attention.  Initial efforts produced ``optimal'' trees based
on network communication models that assumed  
equal \mbox{point-to-point} latencies 
between any two processes.  This assumption is violated in
most practical settings, however, particularly in heterogeneous 
systems such as clusters of SMPs and wide-area ``computational Grids,''
with the result that collective operations 
perform suboptimally.  In response, more recent work has focused 
on creating {\em topology-aware} trees for collective operations that minimize
communication across slower channels (e.g., a wide-area network).  While these 
efforts have significant communication benefits, they all limit their view of 
the network to only two layers.  We present a strategy based upon a 
{\em multilayer} view of the network.  By creating {\em multilevel 
topology-aware} trees we take advantage of communication cost differences 
at every level in the network.  We used this strategy to implement 
topology-aware versions of several MPI collective operations 
in \hbox{MPICH-G2}, the Globus Toolkit$^{TM}$-enabled version of the 
popular MPICH implementation of the MPI standard.  
Using information about topology provided by \hbox{MPICH-G2},
we construct these multilevel 
topology-aware trees automatically during execution.
We present results demonstrating the advantages of our 
multilevel approach by comparing it to the default (topology-unaware)
implementation provided by MPICH and a topology-aware two-layer implementation.
\end{abstract}


\pagebreak

\section{Introduction}

The problem of building ``optimal'' communication trees for
collective operations
has received much attention in recent years.
The telephone model, which assumes that send and receive times are
equal and that messages are not packetized, implies that the optimal broadcast 
algorithm uses a binomial tree.
Under models that expand the telephone model to account for message latency,
such as the postal~\cite{postal} or LogP~\cite{logp} models, the communication 
topology of an optimal
broadcast algorithm becomes a generalized Fibonacci tree.
All of these approaches construct optimal trees for collective operations by first 
modeling the communication characteristics of a network with a set of 
parameters and then building the optimal trees based on parameter values and 
their model.

Underlying this work is
the assumption that the communication times between all process pairs
in the computation are equal.  While this is a reasonable approximation 
when the entire computation is performed on a single machine, 
it is not reasonable when the computation is executed on a cluster of 
symmetric multiprocessors (SMPs) in a local-area network, or worse, in 
a {\em computational Grid}~\cite{Globus,GridBook,globus-physics} environment, 
in which
multiple parallel computers are connected by
local-area, campus-area, or even wide-area networks.  Rapid
improvements in network performance have engendered considerable
interest in parallel computing
in the last context, as evidenced by experiments and
initiatives such as
the I-WAY~\cite{isoftcpe}, National Technology Grid~\cite{CACMgrid}, 
Information Power Grid~\cite{IPG99}, and TeraGrid~\cite{teragrid}.

Under these circumstances the trees produced by the conventional
models perform suboptimally.
In such heterogeneous environments, communication costs over different links
can differ by an order of magnitude or
more.  In these situations, {\em topology-aware} algorithms can 
dramatically
improveme the performance.  For example, in the case of N
processors distributed into two clusters, a traditional
reduction algorithm may generate \mbox{O(log N)} intercluster messages,
while a topology-aware algorithm generates only 1, for a
cost saving of a factor of \mbox{O(log N)} if intercluster message 
costs dominate.

Previous work~\cite{starT,magpie} has demonstrated that 
topology-aware 
collective operations can indeed reduce communication costs by reducing 
the amount of communication performed over slow channels.  However, this work
limited the depth of network stratification to only two levels: other 
processors are either near or far.  In~\cite{optcollops} we compared a
prototype of our multilevel approach to the topology-{\em un}aware binomial
tree algorithm distributed with MPICH and to MagPIe, one of the topology-aware
two-level techniques.  In that prototype we ``guessed'' which computers shared
a local network by inspecting their fully qualified domain names, and 
thereafter representing our multilevel clustering of processes with a
sequence of {\em hidden communicators} inside MPI communicators.

In this paper we present a much improved refinement of that prototype 
that allows collective operations to exploit knowledge concerning the 
structure of a multilevel network, in which the neighbors are processors
that are categorized according to their expected \mbox{point-to-point}
communication characteristics.  The identification of which processes
share a local network is now a simple matter of users providing values
for selected environment variables.  Additionally the use of hidden
communicators to represent the multilevel clustering has been replaced
by integer vectors.  The use of hidden communicators required us to
implement the collective operations as a {\em sequence of collective
operations}, for example, an \verb+MPI_Bcast+ was implemented as a 
sequence of \verb+MPI_Bcast+s sequencing over each of the hidden communicators
in turn, which typically resulted in the use of binomial trees at each level.  
By replacing the hidden communicators with integer vectors we
are now free to implement collective operations using \mbox{point-to-point}
operations over any tree we create.

To permit experimental studies, we have implemented our multilevel 
approach for five of the collective operations supported by the Message Passing
Interface (MPI) standard~\cite{mpi-forum:journal}: \verb+MPI_Bcast+, 
\verb+MPI_Reduce+, \verb+MPI_Barrier+, \verb+MPI_Gather+, and 
\verb+MPI_Scatter+.  We use \hbox{MPICH-G2}~\cite{mpichg2}, the successor to 
\hbox{MPICH-G}~\cite{mpi-nexus-pc}, which 
is based on the popular MPICH implementation~\cite{mpich} of the
MPI standard. \hbox{MPICH-G2} uses services provided 
by the Globus Toolkit$^{TM}$, or simply Globus, to support execution in 
heterogeneous and distributed 
environments.  This use of \hbox{MPICH-G2} enables experimentation within
realistic wide-area environments that would not otherwise be easily accessible.

In the sections that follow, we describe our multilevel topology approach.
Then, we present experimental results that illustrate the benefits of
our multilevel approach by comparing it with (1) the topology-{\em un}aware
implementation currently distributed with MPICH and (2) MagPIe~\cite{magpie},
one of the topology-aware two-level implementations of collective operations.
We briefly discuss other recent topology-aware and optimized collective 
operations efforts and conclude with a discussion of future work.

\section{Multilevel Topology-Aware Approach}

\begin{figure}
\begin{centering}
\includegraphics{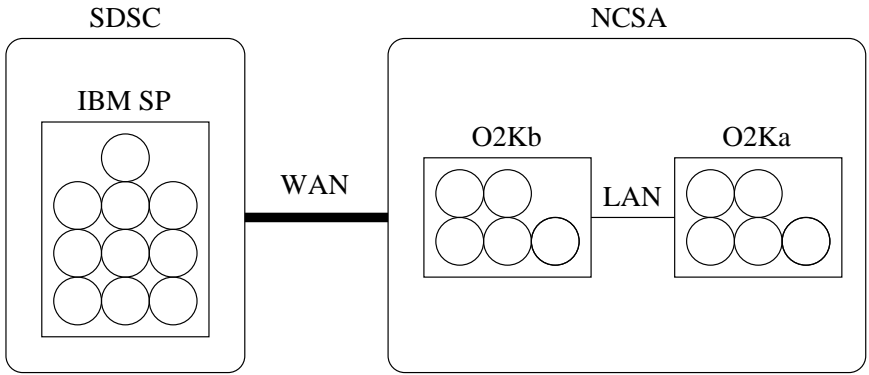}
\caption{An example of a Grid computation involving 10 processes on one 
    IBM~SP at SDSC and another 10 processes distributed evenly across two 
    SGI~Origin2000s (O2K$_a$ and O2K$_b$) at NCSA.}\label{fig-mach}
\end{centering}
\end{figure}

Figure~\ref{fig-mach} depicts an MPI application
involving 20 processes distributed over three machines located at
the San Diego Supercomputer Center (SDSC) and the National Center
for Supercomputing Applications (NCSA).
We depict 10 processes on the IBM~SP at SDSC and 5 processes
on each of two Origin2000s, O2K$_a$ and O2K$_b$, at NCSA.  
The slowest communication is between sites, which uses TCP over a
wide-area network, with faster communication between the O2Ks at NCSA,
which uses TCP over their local-area network, and the fastest
communication, of course, within each machine.  

In the remainder of this section we describe a broadcast using first
the topology-unaware implementation currently distributed with MPICH, 
then a 2-level topology-aware approach, and finally our multilevel
topology-aware broadcast.

\subsection{A Topology-{\em Un}aware Broadcast}

\begin{figure}
\begin{centering}
\includegraphics{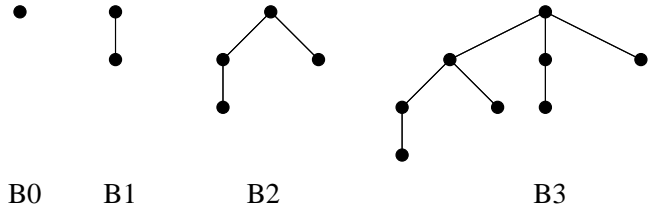}
\caption{The binomial trees $B_0$ through $B_3$.}\label{fig-binomial}
\end{centering}
\end{figure}

Topology-unaware implementations of broadcast, including the one
distributed with MPICH, often make the simplifying 
assumption that the communication times between all process 
pairs in the computation are equal.  Under this assumption the broadcast
is often implemented by using a {\em binomial tree}.

A binomial tree $B_k$ is an ordered tree (i.e., children of each node
are ordered) of order $k \ge 0$ defined recursively.  As 
shown in Figure~\ref{fig-binomial}, the binomial tree $B_0$ consists
of a single node.  The binomial tree $B_k$ ($k > 0$) has a root
with $k$ children where the $i^{th}$ child ($0 < i \le k$) is the root 
of the binomial tree $B_{k-i}$.  Figure~\ref{fig-binomial} depicts the binomial 
trees $B_0$ through $B_3$.

When communication times between all process pairs in the computation are
equal and have relatively low latency, \hbox{Bar-Noy} and Kipnis 
show that implementing a broadcast with a binomial tree has the desirable 
property that all processes will complete the broadcast at approximately
the same time thus, achieving proper load balancing~\cite{postal}.

\subsection{A 2-Level Topology-Aware Broadcast}
\label{subsec-2lvl}

\begin{figure}
\begin{centering}
\includegraphics{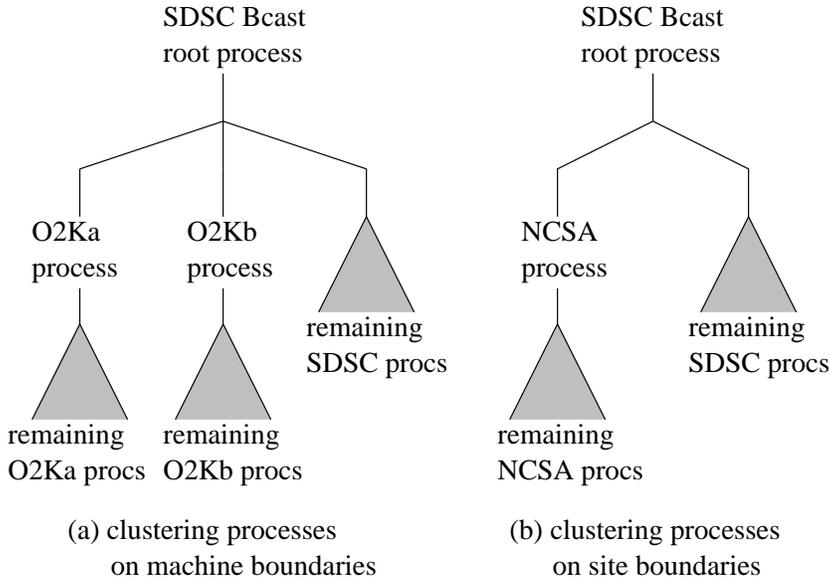}
\caption{An example of two 2-level topology-aware broadcast trees
    rooted at SDSC spanning 2 Origin2000s (O2K$_a$ and O2K$_b$) at NCSA 
    and an IBM~SP at SDSC: (a) clustering processes on {\em machine boundaries} 
    and (b) clustering on {\em site boundaries}.}\label{fig-2lvlbcast}
\end{centering}
\end{figure}

Existing 2-level topology-aware approaches~\cite{starT,magpie} cluster 
processes into groups.  The two natural choices for the machines
depicted in Figure~\ref{fig-mach} are to cluster the processes based
either on {\em machine boundaries}, creating three groups -- the IBM~SP, O2K$_a$, 
and O2K$_b$, or {\em site boundaries} creating two groups -- SDSC and NCSA.  
While both are reasonable choices and would improve
performance when compared with the topology-{\em un}aware binomial tree
distributed with MPICH,
both choices ignore the disparity in network performance between the
local- and wide-area networks.  Consider, for example,
a broadcast rooted at one of the processes at SDSC.  Figure~\ref{fig-2lvlbcast}a
depicts the broadcast tree of the 2-level approach when the processes
are clustered on machine boundaries.  The broadcast starts
with the SDSC root process sending messages to designated processes on each 
of the O2Ks at NCSA, resulting in two messages travelling across the wide-area 
network, and concludes with broadcasts within each machine.  By contrast, 
Figure~\ref{fig-2lvlbcast}b depicts the broadcast tree when the processes
are clustered on site boundaries.  In this case the root at SDSC
sends a single message across the wide-area network to a process on 
one of the two O2Ks at NCSA and concludes with a broadcast within the 
IBM~SP with another simultaneous broadcast across all the processes
at NCSA, which would typically require multiple messages to travel
across NCSA's local network.

\subsection{A Multilevel Topology-Aware Broadcast}

\begin{figure}
\begin{centering}
\includegraphics{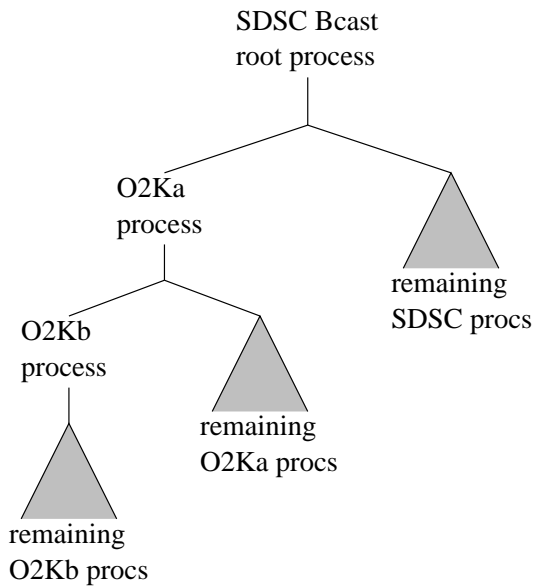}
\caption{An example of a multilevel topology-aware broadcast tree rooted
    at SDSC spanning 2 Origin 2000s (O2K$_a$ and O2K$_b$) at NCSA and an 
    IBM~SP at SDSC.}\label{fig-mlvlbcast}
\end{centering}
\end{figure}

The multilevel topology-aware approach we present minimizes messaging 
across the slowest links {\em at each level} by clustering the processes 
at the wide-area level into site groups, and then within each site group, 
clustering processes at the local-area level into machine groups.
Using the same broadcast example from Section~\ref{subsec-2lvl}, we depict in
Figure~\ref{fig-mlvlbcast} the broadcast tree used by a multilevel approach.
Here the broadcast starts with the SDSC root process sending a single message 
across the wide-area network to one of the processes at NCSA, in 
Figure~\ref{fig-mlvlbcast} we depict a process on O2K$_a$.  The broadcast 
continues with the receiving process on O2K$_a$ sending a single message 
across NCSA's local network to a process on O2K$_b$ and the entire broadcast 
concludes with broadcasts within each machine.  This multilevel 
clustering minimizes messaging over the slower wide- and local-area 
networks.

\section{Multilevel Topology-aware Approach in \hbox{MPICH-G2}}

In this section we describe our implementation of multilevel topology-aware
collective operations in the Globus Toolkit-based \hbox{MPICH-G2}.  For illustrative purposes,
we discuss our implementation of \verb+MPI_Bcast+ in detail.

\subsection{RSL Specification of Topology}

MPICH-G2 uses the Globus Toolkit's Resource Specification Language~(RSL)~\cite{GRAM97} to
describe the resources required to run an application.  
Users write {\em RSL scripts}, which identify resources 
(e.g., computers) and specify requirements (e.g., number of CPUs, memory, 
execution time) and parameters (e.g., location of executables, command 
line arguments, environment variables) for each.  An RSL script can
be used as
the user interface to \verb+globusrun+, an upper-level Globus service
that first authenticates the user by using the Grid Security 
Infrastructure~(GSI)~\cite{GSI-journal} and then
schedules and monitors the job across the various machines by using
two other Globus Toolkit services: the Dynamically-Updated Request Online 
Coallocator~(DUROC)~\cite{CoAllocation99} and Grid Resource Allocation 
and Management~(GRAM)~\cite{GRAM97}.  RSL is designed to be an
easy-to-use language to describe multiresource multisite jobs while
hiding all the site-specific details associated with requesting
such resources.

\begin{figure}
\begin{footnotesize}
\begin{verbatim}
+
( &(resourceManagerContact="sp.npaci.edu") 
   (count=10)
   (jobtype=mpi)
   (label="subjob 0")
   (environment=(GLOBUS_DUROC_SUBJOB_INDEX 0))
   (directory=/homes/users/smith)
   (executable=/homes/users/smith/myapp)
)
( &(resourceManagerContact="o2ka.ncsa.uiuc.edu") 
   (count=5)
   (jobtype=mpi)
   (label="subjob 1")
   (environment=(GLOBUS_DUROC_SUBJOB_INDEX 1))
   (directory=/users/smith)
   (executable=/users/smith/myapp)
)
( &(resourceManagerContact="o2kb.ncsa.uiuc.edu") 
   (count=5)
   (jobtype=mpi)
   (label="subjob 2")
   (environment=(GLOBUS_DUROC_SUBJOB_INDEX 2))
   (directory=/users/smith)
   (executable=/users/smith/myapp)
)
\end{verbatim}
\end{footnotesize}
\caption{\strut{An RSL script for an \hbox{MPICH-G2} application running
    on three machines that facilitates {\em 2-level process 
    clustering}.}\label{fig-rsl2lvl}}
\end{figure}

\begin{figure}
\begin{footnotesize}
\begin{verbatim}
+
( &(resourceManagerContact="sp.npaci.edu") 
   (count=10)
   (jobtype=mpi)
   (label="subjob 0")
   (environment=(GLOBUS_DUROC_SUBJOB_INDEX 0))
   (directory=/homes/users/smith)
   (executable=/homes/users/smith/myapp)
)
( &(resourceManagerContact="o2ka.ncsa.uiuc.edu") 
   (count=5)
   (jobtype=mpi)
   (label="subjob 1")
   (environment=(GLOBUS_DUROC_SUBJOB_INDEX 1)
       (GLOBUS_LAN_ID NCSAlan))
   (directory=/users/smith)
   (executable=/users/smith/myapp)
)
( &(resourceManagerContact="o2kb.ncsa.uiuc.edu") 
   (count=5)
   (jobtype=mpi)
   (label="subjob 2")
   (environment=(GLOBUS_DUROC_SUBJOB_INDEX 2)
       (GLOBUS_LAN_ID NCSAlan))
   (directory=/users/smith)
   (executable=/users/smith/myapp)
)
\end{verbatim}
\end{footnotesize}
\caption{\strut{An RSL script for an \hbox{MPICH-G2} application running
    on three machines that facilitates {\em multilevel process 
    clustering}.}\label{fig-rslMlvl}}
\end{figure}

Figure~\ref{fig-rsl2lvl} depicts an RSL script for an \hbox{MPICH-G2}
application intended to run on the computational Grid depicted in
Figure~\ref{fig-mach}.  It depicts a job as a set of three {\em subjobs}, 
where each subjob is associated with a particular resource, in our example, 
a computer.  Subjobs define a natural machine-boundary partitioning of the
processes in \verb+MPI_COMM_WORLD+ and are sufficient for a 2-level machine
boundary clustering of the processes.  To achieve a multilevel clustering,
the user must identify those machines that are on the same local network
by specifying a value for an \hbox{MPICH-G2}-defined environment variable 
\verb+GLOBUS_LAN_ID+, as depicted in the RSL script in 
Figure~\ref{fig-rslMlvl}.  Specifying the same value (\verb+NCSAlan+) 
in the second and third subjobs instructs \hbox{MPICH-G2} to cluster these 
two machines into the same local-area network group.  This same technique 
can be used to cluster many subjobs in the same local-area network group 
while simultaneously creating multiple local-area network groups through 
the assignment of multiple yet unique \verb+GLOBUS_LAN_ID+ values.  This 
simple specification (the only difference between Figures~\ref{fig-rsl2lvl} 
and~~\ref{fig-rslMlvl}) is all that is required to create multilevel 
topology-aware clustering of the processes.

The multilevel clustering information specified in RSL (i.e., processes
gathered first into machine groups and then local network groups composed
of machine groups) creates a multilevel grouping of the processes in 
\verb+MPI_COMM_WORLD+ and is distributed to all the processes during 
\hbox{MPICH-G2} bootstrapping to be stored within \verb+MPI_COMM_WORLD+ on
each process.  When new communicators are created (e.g., via \verb+MPI_Comm_split+),
\hbox{MPICH-G2} propagates the relevant multilevel clustering information to 
the newly created communicator so that {\em all communicators} in 
\hbox{MPICH-G2} have the multilevel clustering information pertaining to their 
process groups.  As an interesting side effect we have made this multilevel 
topology information available to MPI applications through existing
MPI communicator caching idioms. See~\cite{mpichg2} for a full 
description of \hbox{MPICH-G2}'s topology discovery mechanism.
\subsection{MPICH-G2's Multilevel Topology-Aware Broadcast}

A multilevel topology-aware clustering of processes is not sufficient
in itself to allow the construction of a broadcast tree such as that depicted in 
Figure~\ref{fig-mlvlbcast}: \hbox{MPICH-G2} also needs to know which process is 
the root of the broadcast.  Construction of the multilevel 
topology-aware tree is therefore deferred until the application calls a 
collective operation.  At that time each process simultaneously and 
independently (i.e., without communication) construct an identical tree based 
on the multilevel process grouping found in the communicator and the parameters 
passed (e.g., identifying the root process of a broadcast) to the collective 
operation.

One benefit of using a multilevel topology-aware tree to implement 
a collective operation is that we are free to select different subtree 
topologies at each level.  For example, a multilevel broadcast tree can start
with a broadcast from the root to selected processes at each site across a 
wide-area network, followed by broadcasts at each site to selected processes 
on each machine across the local networks, and concluding with broadcasts 
within each machine.  We have the freedom to use different broadcast 
topologies at each stage in the sequence.  Bar-Noy and Kipnis show
that in high-latency networks (e.g., a wide-area network) 
the optimal broadcast topology is a flat tree in which the root sends the 
data to all other processes directly, while in a low-latency network 
(e.g., intramachine messaging), the optimal broadcast topology is a binomial 
tree~\cite{postal}.  We take advantage of these findings and the flexibility 
of our multilevel approach in our implementation of \verb+MPI_Bcast+ by using
a flat broadcast tree at the initial wide-area level and binomial trees at 
the local-area and intramachine levels.

In the next section we present results demonstrating the advantages of our
multilevel approach by comparing it with the default (topology-{\em un}aware)
implementation provided by MPICH and a topology-aware two-layer implementation.

\section{Experimental Results}
\label{sec-exp}

To demonstrate the advantages of our multilevel approach, we examine
its effects on \verb+MPI_Bcast+.  
The MPICH implementation of \verb+MPI_Bcast+ is based on binomial trees;
hence, in a distributed heterogeneous environment like a computational
Grid its performance is acutely sensitive to the distribution 
of the processes and the root of the broadcast.  
For example, in an application using $P=2^k$ processes distributed evenly 
across $C=2^i, 0 \le i \le k$ clusters, a broadcast implemented using
a binomial tree propagates the message down its longest path 
using at least $log_2C$ intercluster messages and $log_2\frac{P}{C}$
intracluster messages.
In contrast, 
under certain intercluster network performance
conditions described by \mbox{Bar-Noy} and Kipnis in their postal model,
our multilevel method could be used to send 
1 intercluster message and $log_2\frac{P}{C}$ intracluster messages.
Assuming an intercluster latency $l_s$ sec and bandwidth $b_s$ Kb/sec; 
and an intracluster latency $l_f$ sec and bandwidth $b_f$ Kb/sec, 
broadcasting a message of N Kb using the 
binomial tree conservatively takes 
\mbox{$O((logC)(l_s+\frac{N}{b_s}) + (log\frac{P}{C})(l_f+\frac{N}{b_f}))$},
whereas
broadcasting the same message using our multilevel method takes only 
\mbox{$O((l_s+\frac{N}{b_s}) + (log\frac{P}{C})(l_f+\frac{N}{b_f}))$}.

\begin{figure}
\begin{small}
\begin{verbatim}
For (each message size M)
  MPI_Barrier(MPI_COMM_WORLD)
  if (MPI_COMM_WORLD rank == 0)
    t0 = get_time()
  For (r = 0; r < Nprocs; r ++)
    MPI_Bcast(root=r to MPI_COMM_WORLD message size M)
    ack_barrier()
  if (MPI_COMM_WORLD rank == 0)
    t1 = get_time()
    report message size M, time t1-t0
\end{verbatim}
\end{small}
\caption{\strut{The broadcast timing application.}\label{fig-tim-bcast}}
\end{figure}

We wrote a small MPI application (depicted in Figure~\ref{fig-tim-bcast}) 
that times the broadcasts of messages of increasing size.  
To represent a broadcast with an arbitrary
root, we timed how long it would take to broadcast each message
of size M as each process in \verb+MPI_COMM_WORLD+ took its turn
as the root.  Also, in order to eliminate any potential pipelining
that might occur between consecutive broadcasts, we inserted a barrier
(\verb+ack_barrier()+) after each broadcast in which all processes
other than rank 0 \verb+MPI_Send+ an ACK message to process 0 and then
wait to \verb+MPI_Recv+ a GO message.  Process 0, after \verb+MPI_Recv+'ing
the ACK message from all the other processes, \verb+MPI_Send+'s a GO
message to each of the other processes, one at a time.  We chose to write our
own barrier rather than calling \verb+MPI_Barrier+ because 
we have reimplemented \verb+MPI_Barrier+ to reflect multilevel topology and
we wished these tests to reflect the differences only in the broadcast 
implementations.

\begin{figure}
\begin{centering}
\includegraphics{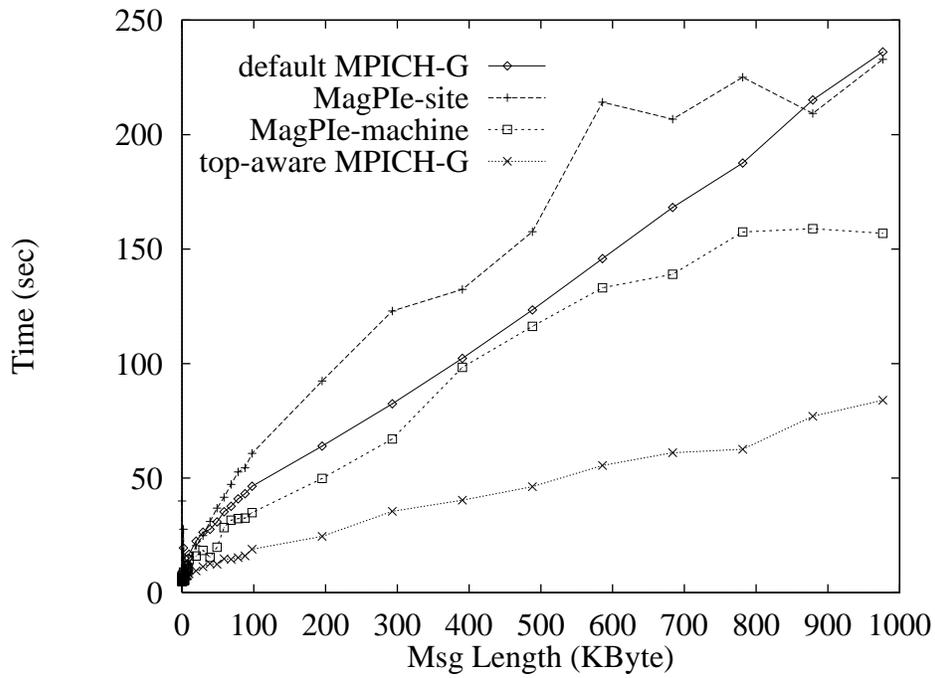}
\caption{Original MPICH broadcast~vs.~topology-aware MPICH broadcast~vs.~MagPIe
    broadcast running 16 processes 
    on the IBM~SP at SDSC and 16 processes on each the IBM~SP and 
    SGI~Origin2000 at ANL.}\label{fig-bcast-magpie}
\end{centering}
\end{figure}

We conducted experiments running the MPI
application depicted in Figure~\ref{fig-tim-bcast} on 
three computers: the IBM~SP at the San Diego Supercomputer Center
(SDSC-SP) and the IBM~SP (ANL-SP) and SGI~Origin200 (ANL-O2K) at
Argonne National Laboratory.
We compare
our multilevel topology approach to the binomial tree provided by MPICH
and include comparisons to the 2-level approach provided by MagPIe.  
We ran the application four times, each time using 
16 processes on each of the three computers.  These results are
depicted in Figure~\ref{fig-bcast-magpie}.  
The curves labeled ``MagPIe-machine'' and ``MagPIe-site'' represent
two runs using MagPIe version 2.0.1, each time with 
a different cluster definition.  
In our first MagPIe 
run (``MagPIe-machine'') we defined three clusters, one for each computer, 
of 16 processes each.  In our second MagPIe run (``MagPIe-site'') we defined 
two clusters: an ANL cluster comprising the two ANL machines having 32 
processes and an SDSC cluster comprising the SDSC-SP having only 16 processes.

Figure~\ref{fig-bcast-magpie} 
shows there are significant benefits to the multilevel approach when
compared with a simple binomial tree and even when compared with a 2-level
approach as implemented by MagPIe.
A multilevel view of the network allows an application to avoid 
slower channels {\em at each level}.  In our experiments, the broadcast is 
optimized by sending one message across the wide-area network, then one 
message across the local-area network, and then many messages
within each computer.

\section{Related Work}

Previous efforts have focused on creating ``optimal'' trees for collective 
operations where \mbox{point-to-point} communications are not necessarily 
equal between any two processes.  Husbands and Hoe present 
MPI-StarT~\cite{starT}, an MPI implementation for a cluster of SMPs 
interconnected by a high-performance interconnect.  They report significant 
improvements after modifying the MPICH broadcast algorithm, which uses 
binomial trees.  Their modifications use information that describes their 
cluster topology by minimizing intercluster communication during collective 
operations. MagPIe~\cite{magpie} is another MPI system designed to construct 
collective operation trees in heterogeneous communication environments.  
MagPIe recognizes a two-layer communication network that distinguishes between 
local- and wide-area communication.  By minimizing wide-area communication, 
much in the same way MPI-StarT minimizes intercluster communication, MagPIe 
has seen significant improvements in all the MPI collective operations.

Both efforts have produced impressive results and clearly demonstrate 
that there are significant advantages to implementing collective operations in
a topology-aware manner.  However, both limit
their view of the network to only two layers; MPI-StarT distinguishes
between intra- and intercluster communication within the same
local-area, and MagPIe distinguishes between local- and wide-area communication.
There are opportunities for further optimization by 
using trees that stratify the network deeper than two layers.

In~\cite{pipelining} van de Geijn et al. show the advantages of
implementing collective operations by segmenting and pipelining
messages when communicating over relatively slower channels (e.g., TCP
over local- and wide-area networks).

In~\cite{magpie-PC} Kielman et al. extend MagPIe by incorporating
van de Geijn's pipelining idea through a technique they call Parameterized
LogP (PLogP), which is an extension of the LogP model presented 
by Culler et al~\cite{logp}.
In this extension, MagPIe still recognizes only a two-layer communication
network, but through parameterized studies of the network, the researchers determine
``optimal'' packet sizes.  This technique works well for applications
that always run on the same computational grid having relatively 
stable performance, but requires retuning when moving the application
from one computing environment or network to another.

\section{Future Work}

We have implemented five of the MPI collective operations in a 
topology-aware multilevel manner in \hbox{MPICH-G2}.  Encouraged by our
initial results, we plan to upgrade \hbox{MPICH-G2's} remaining MPI collective 
operations in a similar manner.

Our general strategy implements a collective operation by first stratifying 
the network into multiple levels and then minimizing the communication
across the slowest channels.  In doing so, however, we may encounter
a tree that has multiple siblings at a particular level, for example,
many sites connected across the wide-area network or many machines
at a particular site.  When this situation happens, we implement the collective
operation at that level using a binomial tree at all but the wide-area
network level.  Unfortunately, a binomial tree is not always the best choice.
\mbox{Bar-Noy} and Kipnis show that the shape of a collective 
operation tree depends heavily on the \mbox{point-to-point} communication 
characteristics of the send/receive primitives on which it is implemented.  
Their model incorporates a latency parameter $\lambda \ge 1$.  They show
that for low latencies, (for example, communication within a single machine), 
the optimal broadcast tree is a binomial tree, but for higher latencies,
(for example, communication across a wide-area network), the optimal broadcast
tree becomes flatter.  We will investigate ways to select better, if
not optimal, collective operation trees by choosing those that respect
the different communication characteristics at each level of our multilevel
view.

The pipelining techniques presented by van de Geijn et al. can be used at 
each of the levels in \hbox{MPICH-G2's} multilevel topology-aware collective 
operations.  Using techniques similar to Kielman's PLogP method, we will 
develop methods to determine the appropriate packet sizes with respect to 
network performance at {\em each level} of our multilevel view.

\section{Summary}

As Grid computations become increasingly prevalent, the need for 
topology-aware collective operations also increases.  We have a version 
of \hbox{MPICH-G2} that implements five collective operations in a multilevel 
topology-aware manner.  We have shown, at least for \verb+MPI_Bcast+, that when 
compared with the binomial tree provided by MPICH and the 2-level approach 
provided by MagPIe there are significant advantages to executing 
collective operations using a multilevel view of the network.
Through a simple process of identifying machines that are common
to a local-area network, we have provided a means by which an MPI
application may take advantage of the multilevel topology-aware algorithms
without requiring code modifications or special functions.

\begin{acknowledge}

We thank the San Diego Supercomputer Center and the National Center
for Supercomputing Applications for providing access to their
machines.  We also thank the members of the Globus development
team for their support, patience, and many ideas. This work was
supported in part by the Mathematical, Information, and Computational
Sciences Division subprogram of the Office of Advanced Scientific
Computing Research, U.S. Department of Energy, under Contract
W-31-109-Eng-38; by the U.S. Department of Energy under
Cooperative Agreement No. DE-FC02-99ER25398; 
by the National Science Foundation; by DARPA; and by
the NASA Information Power Grid program.

\end{acknowledge}

\bibliographystyle{plain}
\bibliography{foster_bibliography,mpi-allbib,mpi-book,globus,prop}

Nicholas T. Karonis received a B.S. in finance and a B.S. in computer
science from Northern Illinois University in 1985, an M.S. in computer
science from Northern Illinois University in 1987, and a Ph.D. in
computer science from Syracuse University in 1992.  He spent
summers from 1981 to 1991 as a student at Argonne National Laboratory
where he worked on the p4 message-passing library, automated reasoning,
and genetic sequence allignment.  From 1991 to 1995 he worked on the control
system at Argonne's Advanced Photon Source and from 1995 to 1996
for the Computing Division at Fermi National Accelerator Laboratory.
Since 1996 he has been an assistant professor of computer science at
Northern Illinois University and a resident associate guest of Argonne's
Mathematics and Computer Science Division where he has been a member
of the Globus Project.  His current research interest is message-passing 
systems in computational Grids.

Bronis R. de Supinski is a computer scientist in the Center
for Applied Scientific Computing at Lawrence Livermore
National Laboratory. His research interests include
message passing implementations and tools, memory performance
improvement, cache coherence and distributed shared memory,
consistency semantics and performance evaluation modeling
and tools. Bronis earned his Ph.D. in computer science
from the University of Virginia in 1998. He is a
member of the ACM and the IEEE Computer Society.

Ian Foster received his B.Sc. (Hons I) at the University of Canterbury 
in 1979 and his Ph.D. from Imperial College, London, in 1998. He
is senior scientist and associate director of the
Mathematics and Computer Science Division at Argonne National
Laboratory, and professor of computer science at the University of
Chicago.  He has published four books and over 150 papers and
technical reports.  He co-leads the Globus Project, which provides
protocols and services used by industrial and academic distributed
computing projects worldwide.  He co-founded the influential Global
Grid Forum and co-edited the book ``The Grid: Blueprint for a New
Computing Infrastructure.''

William Gropp received his B.S. in mathematics from Case Western Reserve 
University in 1977, a an M.S. in physics from the University of Washington in 
1978, and a Ph.D. in computer science from Stanford in 1982. He held the 
positions of assistant (1982-1988) and associate (1988-1990) professor in 
the Computer Science Department at Yale University. In 1990, he joined the 
numerical analysis group at Argonne, where he is a senior computer 
scientist and associate director of the Mathematics and Computer Science 
Division, a senior scientist in the Department of Computer Science at the 
University of Chicago, and a Senior Fellow in the Argonne-University of Chicago 
Computation Institute. His research interests are in parallel computing, 
software for scientific computing, and numerical methods for partial 
differential equations. He has played a major role in the development of 
the MPI message-passing standard. He is co-author of MPICH, the most widely used 
implementation of MPI, and was involved in the MPI Forum as a 
chapter author for both MPI-1 and MPI-2. He has written many books and 
papers on MPI including "Using MPI" and "Using MPI-2". He is also one of 
the designers of the PETSc parallel numerical library, and has developed 
efficient and scalable parallel algorithms for the solution of linear and 
nonlinear equations.

Ewing Lusk received his B.A. in mathematics from the University of Notre Dame
in 1965 and his Ph.D. in mathematics from the University of Maryland in 1970.
He is currently a senior computer scientist in the Mathematics and Computer
Science Division at Argonne National Laboratory.  His current projects include
implementation of the MPI message-passing standard, research into programming
models for parallel architectures, and parallel performance analysis tools.
He is a leading member of the team responsible for MPICH implementation of the
MPI message-passing interface standard.  He is the author of five books and
more than seventy-five research articles in mathematics, automated deduction,
and parallel computing.

Sebastien Lacour graduated in physics in 1999 at the Ecole
Normale Superieure of Lyon, France. He received his master's
degree in computer science in 2002 at IFSIC, University of
Rennes, France.  He is currently a Ph.D. student at IRISA/INRIA in
Rennes.  His research interests include networks, compilation, and
parallel and distributed systems.  His current work focuses on
distributed shared-memory systems over large-scale, hierarchical
architectures (multicluster platforms).

\end{document}